\documentclass[12pt]{article}
\usepackage{hyperref,amsfonts,amssymb}

\def\C{\mathbb{C}}

\def\R{\mathbb{R}}

\def\vk{\varkappa}

\def\dsize{\displaystyle}

\def\nn{ \nonumber }

\def\defeq {\stackrel{\mbox{\rm\small def}}{=}}

\def\bq{ \begin{equation} }
\def\eq{ \end{equation} }
\def\ben{ \begin{eqnarray} }
\def\en{ \end{eqnarray} }

\def\frac#1#2{{#1\over #2}}
\def\dfrac#1#2{{\displaystyle{#1\over#2}}}
\def\on#1#2{\mathop{\vbox{\ialign{##\crcr\noalign{\kern2pt}
$\scriptstyle{#2}$\crcr\noalign{\kern2pt\nointerlineskip}
\kern-2pt$\hfil\displaystyle{#1}\hfil$\crcr}}}\limits}

\newtheorem{prop}{Proposition}
\newtheorem{theor}{Theorem}

\begin{document}

\baselineskip=15pt
\vspace{1cm} \noindent {\LARGE \textbf{Poisson maps and integrable
deformations\\ of the Kowalevski top }} \vskip1cm \hfill
\begin{minipage}{13.5cm}
\baselineskip=15pt
{\bf
 I V Komarov ${}^{1}$,
 V V Sokolov ${}^{2}$ and
 A V Tsiganov ${}^{1}$ }\\ [2ex]
{\footnotesize
${}^1$ V.A. Fock Institute of Physics, St.Petersburg State University,
St.Petersburg, Russia\\
${}^2$ Landau Institute for Theoretical Physics, Moscow, Russia}
\\
\vskip1cm{\bf Abstract}

We construct a Poisson map between manifolds with linear Poisson brackets
corresponding to the Lie algebras $e(3)$ and $so(4)$. Using this map
we establish a connection between the  deformed  Kowalevski top on $e(3)$
proposed by Sokolov and the Kowalevski top on $so(4)$. The connection between
these systems leads to the separation of variables for the deformed system on $e(3)$
and yields the natural $5\times 5$ Lax pair for the Kowalevski top on
$so(4)$.
\end{minipage}

\vglue1cm \textbf{Corresponding Author}:
I V Komarov, Institute of Physics, St.Petersburg State University,
St.Petersburg, Russia
E-mail: \textsc{%
igor.komarov@pobox.phys.spbu.ru\/} \newpage

\section{Introduction}
In 1888 Sophie Kowalevski \cite{kow89} found and integrated new integrable case of
rotation of a heavy rigid body around a fixed point. In modern terms, this is an
integrable system on the orbits of the Euclidean Lie algebra $e(3)$ with a
quadratic and a quartic in angular momenta integrals of motion.

The Kowalevski top can be generalized in several directions. We can
change either initial phase space or  the form of the
Hamilton function. In 1981 the first author has considered the
Kowalevski top on $so(4)$, $e(3)$  and $so(3,1)$  Lie algebras \cite{kom}.
Separation of variables for these generalizations  was
constructed in \cite{komkuz}. Recently in 2001, the second author has found
integrable deformations of the Kowalevski Hamiltonian on $e(3)$  and $so(4)$
algebra \cite{soktmf,sokO4,BMS}. A Lax representation for the deformed
Kowalevski Hamiltonian on $e(3)$ was found in \cite{sokts1}.

In this paper we establish an explicit nonlinear map of the
Kowalevski top on $so(4)$ to the deformed Kowalevski case on
$e(3)$. The connection between the systems leads to the separation
of variables for the system on $e(3)$\footnote{
Actually, we had
found first a separation of variables for this model and after
that comparing it with known separation of variables for the
$so(4)$ Kowalevski top found the Poisson map between  $e(3)$ and
$so(4)$.}
 and yields a natural $5\times 5$ Lax pair for the
Kowalevski top on $so(4)$ which was unknown. This Lax matrix
provides an algebraic curve for the Kowalevski top on $so(4)$
different from the generalized Kowalevski curve \cite{komkuz}
associated with known separation of variables. For the deformed
Kowalevski Hamiltonian on $so(4)$ neither a separation of
variables nor Lax representation are found yet.

The existence of the Poisson map between $e(3)$ and $so(4)$ allows us to
construct also a new $so(4)$ generalization of the Goryachev-Chaplygin top.

\section{Deformations of the Kowalevski top}
\setcounter{equation}{0}

The rigid body motion about a fixed point under influence of
gravity is described by six dynamical variables: three components
of the angular momentum $\mathbf J=(J_1,J_2,J_3)$ and three
components of the gravity vector $\mathbf x=(x_1,x_2,x_3)$,
everything with respect to a moving orthonormal frame attached to
the body. The invariance  under rotation about the direction of
gravity leads to conservation of  the angular momentum component along
the gravity vector. When its value is fixed the system usually considered
to have only two degrees
of freedom \cite{arn89} such that the Poisson sphere $S^2$ acting
as a reduced configuration space. The reduced phase space may be
identified with coadjoint orbit of Euclidean $e(3)$ algebra with
the Lie-Poisson brackets
\begin{equation}\label{e3}
\,\qquad \bigl\{J_i\,,J_j\,\bigr\}=\varepsilon_{ijk}J_k\,, \qquad
\bigl\{J_i\,,x_j\,\bigr\}=\varepsilon_{ijk}x_k \,,
\qquad \bigl\{x_i\,,x_j\,\bigr\}=0\,,
\end{equation}
where $\varepsilon_{ijk}$ is the totally skew-symmetric tensor. These brackets
have two Casimir functions
\begin{equation}\label{caz0}
A=\mathbf x^2\equiv\sum_{k=1}^3 x_k^2, \qquad
   B=(\mathbf x \cdot \mathbf
J)\equiv\sum_{k=1}^3  x_kJ_k .
\end{equation}
Fixing their values one gets a generic symplectic leaf of $e(3)$
$$
{\mathcal E}_{ab}:\qquad \{{\mathbf x}\,, {\mathbf J}\,:~A=a,~~ B={b}\}\,,
$$
which is a four-dimensional symplectic manifold.

The Hamilton function for the original Kowalevski top is given by
\begin{equation}\label{H-Kow}
H=J_1^2+J_2^2+ 2J_3^2+2c_1 x_1\,,\qquad c_1\in\mathbb C.
\end{equation}
This Hamiltonian and additional integral of motion
\bq\label{K-Kow}
K=\xi\,\cdot \xi^*,
\eq
where
\[
\xi=(J_1 +i J_2)^2-2c_1 (x_1 +i x_2)\,, \qquad \xi^*=(J_1 -i J_2)^2-2c_1 (x_1 -i x_2),
\]
are in the involution and define the moment map whose fibers are Liouville tori in
${\mathcal E}_{ab}$.

The most general known deformation of the Hamiltonian (\ref{H-Kow}) admitting
quadratic and linear terms is defined by the following Hamiltonian
\bq \label{sHam}
\widehat{H}_\vk=J_1^2+J_2^2+2J_3^2+2c_1y_1 +2c_2J_3y_2-c_2^2y_3^2+
2c_3(J_3 +c_2y_2),\qquad c_1,c_2,c_3\in\mathbb C
\eq
(see equation (3.2) in \cite{sokO4}). The corresponding phase space is a
generic orbit of the $so(4)$ Lie algebra with the Poisson brackets
\begin{equation}\label{bundle}
\,\qquad \bigl\{J_i\,,J_j\,\bigr\}=\varepsilon_{ijk}J_k\,, \qquad
\bigl\{J_i\,,y_j\,\bigr\}=\varepsilon_{ijk}y_k \,,\qquad
\bigl\{y_i\,,y_j\,\bigr\}=\varkappa^2\varepsilon_{ijk}J_k.
\end{equation}
Notice that the deformation parameters are not only $c_{2}$ and
$c_{3}$ in (\ref{sHam}) but also $\varkappa$ entering the Lie algebra (\ref{bundle}).
Because physical quantities ${\mathbf y}, {\mathbf J}$ should be
real, $\vk^2$ must be real too and   algebra (\ref{bundle}) is
reduced to its two real forms  $so(4,\R)$ or $so(3,1,\R)$ for
positive and negative $\vk^2$ respectfully. For brevity we will
call it $so(4)$.

Fixing values $a'$ and $b'$ of the Casimir functions
\begin{equation}
\label{center}
A_\vk={\mathbf y}^2 + \vk^2 {\mathbf J}^2, \qquad B_\vk=({\mathbf
y}\cdot{\mathbf J})
\end{equation}
one gets a four-dimensional orbit of  $so(4)$
$$
{\cal O}_{a'b'}:\qquad \{{\mathbf y}\,, {\mathbf J}\,:\quad A_\vk=a',\quad
 B_\vk=b'\}\,,
$$
which is the reduced phase space for the deformed Kowalevski top.

Performing a linear canonical transformation
\[
\begin{array}{lll}
  J_1\to J_1,\qquad & J_2\to c_4\,(J_2+c_2y_3),\qquad & J_3\to
  c_4\,(J_3-c_2y_2), \\
  \\
  y_1\to y_1,\qquad & y_2\to c_4\,(y_2+\varkappa^2 c_2J_3),\qquad &
  y_3\to c_4\,(y_3-\varkappa^2c_2J_2),
\end{array}
\] where $c_4=(1+\varkappa^2 c_2^2)^{-\frac{1}{2}}$, we  reduce the Hamiltonian
$\widetilde{H}_\vk$ (\ref{sHam}) to the following Hamilton function
\bq \label{sHam2}
\widehat{H}_\vk=J_1^2+(1-\varkappa^2c_2^2)J_2^2+2J_3^2
+2c_1y_1+2c_2(y_2J_3-y_3J_2)+2c_3c_4^{-1}J_3\,,
\eq
which is linear in ${\bf y}$.

Parameter $c_3$ in (\ref{sHam}) and (\ref{sHam2})
corresponds to the Kowalevski gyrostat  \cite{gyro,sokO4}). In this paper
we consider the case $c_3=0$.

The integration procedure for Hamiltonian (\ref{H-Kow}) proposed by
Kowalevski is based on the fact that the additional integral of motion
(\ref{K-Kow}) is a product of two quadratic factors.

For the deformed Kowalevski top (\ref{sHam2}) the second integral of motion
$\widehat{K}_\vk$ can be written as
\bq \label{dInt} \widehat{K}_\vk=\widehat{\xi}\cdot\widehat{\xi}^* +
4\varkappa^2(\mathbf
J^2-J_2^2)\Bigl(c_1^2+c_2^2(\mathbf J^2-J_2^2)\Bigr),
\eq
where
\bq
\begin{array}{c}
\widehat{\xi}=\xi-c_2\{\mathbf J^2,y_1+iy_2\}-c_2^2\left({A}_\varkappa
-\varkappa^2 J_2^2\right)\\[3mm]
\widehat{\xi}^*=\xi^*-c_2\{\mathbf J^2,y_1-iy_2\}
-c_2^2\left({A}_\varkappa -\varkappa^2 J_2^2\right).
\end{array}
\label{xi}
\eq
For the same integral we have also another useful representation
\[
\widehat{K}_\vk=\xi_\vk\cdot\xi_\vk^* + \vk^2 c_1^2\,(2\widehat{H}_\vk-\vk^2
c_1^2)+c_2f(\mathbf x,\mathbf J)\,,
\]
where
\[
 \xi_{\vk}=\xi+\vk^2c_1^2\,, \qquad
\xi^*_{\vk}=\xi^*+\vk^2c_1^2\,.
\]
and the polynomial $f(\mathbf x,\mathbf J)$ can be easily restored from (\ref{dInt}).

It follows from these formulas that the
additional fourth degree integral of motion can be reduced to the product of two
conjugated polynomials in the following two special cases:
\[\begin{array}{lll}
  1.\quad & c_2=0,\qquad & K_\vk=\xi_\vk\cdot\xi_\vk^*, \\
  \\
  2.\quad & \vk=0,\qquad & \widehat{K}=\widehat{\xi}\cdot\widehat{\xi}^* \\
\end{array} \]
as well as for the original Kowalevski top.
If $c_2=0$ the Hamiltonian for the Kowalevski top on the orbits of the  Lie algebra
$so(4)$ is given by the same formula (\ref{H-Kow}):
\begin{equation}\label{Hk}
H_\vk=J_1^2+J_2^2+ 2J_3^2+2c_1 y_1\,,\qquad c_1\in\mathbb C.
\end{equation}
The additional integral of motion $K_\vk=\xi_\vk\cdot\xi_\vk^*$
was found in \cite{kom}. A Lax pair of the Heine-Horozov \cite{HH}
type and a separation of variables was
constructed in \cite{komkuz}.

In the case $\vk=0$ the deformed Hamiltonian (\ref{sHam2})
\bq \label{sHam3}
\widehat{H}=J_1^2+J_2^2+2J_3^2 +2{c}_1x_1+2{c}_2(x_2J_3-x_3J_2),
\eq
on $e(3)$ has been considered in \cite{soktmf,sokO4}.

A Lax pair with a spectral parameter for the Kowalevski top  had been found in
\cite{rs87}.  Using this Lax representation and the standard finite-band
integration technique, the authors found in \cite{brs89} explicit expressions
for the solutions of the Kowalevski top which are much simpler than the original
formulae of Kowalevski and K\"otter. A Lax pair generalizing the corresponding
result by Reyman and Semenov-{T}ian-{S}hansky was found by Sokolov and Tsiganov
in \cite{sokts1}.

Below  we present nonlinear Poisson maps  between  $e(3)$ and $so(4)$ Poisson
manifolds. This allows us to relate various integrable systems on the different
symplectic manifolds. As an example, we found an explicit mapping of integrable
system (\ref{sHam3}) on $e(3)$ to the Kowalevski top (\ref{Hk}) on $so(4)$. Using
this, we find a Lax pair of Heine-Horozov type \cite{HH} and construct a separation
of variables for the system with Hamiltonian (\ref{sHam3}) on $e(3)$
following \cite{komkuz}. On the other hand, using the results of \cite{sokts1} we
construct a Lax pair for the Kowalevski top on $so(4)$.

\section{Poisson maps of $e(3)$ and $so(4)$ manifolds}
\setcounter{equation}{0}

Let ${\cal M}_1$ be a Poisson manifold  with generators
$x_1,\dots x_n$ and Poisson bracket $\{\,,\}_1$, and ${\cal M}_2$ another
Poisson manifold  with generators $X_1,\dots X_m$ and
Poisson bracket $\{\,,\}_2$. A map $\sigma$ defined by
\begin{equation}\label{tran}
X_i=\Psi_i(\mathbf x),
\qquad i=1,\dots,m
\end{equation}
where $\mathbf x=(x_1,\dots x_n)$, is called {\it  Poisson map} (or Poisson
homomorphism) if $\{\sigma(F),\sigma(G)\}_1=\sigma(\{F,G\}_2)$ for any functions $F$ and
$G$ on ${\cal M}_2.$

\vskip0.3truecm\par\noindent
 {\bf Example:} Let $p_i$ and $x_j, \quad i,j=1, 2, 3$  be
canonical variables on manifold ${\cal M}_1$ with a Poisson bracket
$\{p_{i},x_{j}\}_{1}=\delta_{ij}$, and $J_{i}, x_k$ form  the manifold ${\cal M}_2$
with respect to a Poisson bracket  of $e(3)$ Lie algebra: $\,\quad
\bigl\{J_i\,,J_j\,\bigr\}_2=\varepsilon_{ijk}J_k\,, \quad
\bigl\{J_i\,,x_j\,\bigr\}_2=\varepsilon_{ijk}x_k \,,\quad
\bigl\{x_i\,,x_j\,\bigr\}_2=0\, $ with Casimir elements $x_{i}x_{i}=a$ and
$x_{i}J_{i}=b=0$.
 Then the map $\sigma: ~~\{~,~\}_{1} \to \{~,~\}_{2} $ defined by $J_{i}=\epsilon_{ijk}x_{j}p_{k}$
establishes a Poisson map ${\cal M}_1\to {\cal M}_2$.

If both Poisson brackets $\{\,,\}_1$ and $\{\,,\}_2$ are linear (and therefore are
related to some Lie algebras), linear Poisson maps (\ref{tran}) corresponds to
homomorphisms of these Lie algebras.

If ${\cal M}_1$ coincides with ${\cal M}_2,$ the
Poisson maps are called {\it canonical transformations}.
The problem of complete efficient description of all nonlinear canonical
transformations is unsolvable. The reason is that for any function
$f(\mathbf x)$ the flow defined by ODEs $\mathbf x_{t}=\{f\,,\mathbf x\}_1$
yields a one-parameter group of canonical transformations.
However one can investigate some interesting subgroups of nonlinear
canonical transformations.

In this paper we deal with linear Poisson brackets corresponding to the Lie
algebras $e(3)$ and $so(4)$. For brevity we will use the same
notations  both for the Poisson manifolds and the Lie algebras. In the next section
we consider some special subgroups of nonlinear canonical transformations of $e(3)$.

\subsection{Canonical transformations of $e(3)$}
Consider the Poisson manifold $e(3)$ defined by linear brackets
(\ref{e3}). Linear canonical transformations of $e(3)$ consist of rotations
\bq
\label{rotE3}
\mathbf x\to \alpha\, {\rm U}\,\mathbf x\,,\qquad \mathbf J\to {\rm U}\,\mathbf J\,,
\eq
where $\alpha$ is an arbitrary parameter and $U$ is an orthogonal constant matrix,
and shifts
\bq
\mathbf x\to \mathbf x \,,\qquad \mathbf J\to \mathbf J+ {\rm S}\,\mathbf x\
,\label{shiftE3}
\eq
where ${\rm S}$ is an arbitrary $3\!\times\!3$ skew-symmetric constant matrix.
\par\noindent
\textbf{Example 1:} The composition of the scaling $\mathbf x\to\alpha\mathbf x$ and
the rotation around third axis defined by
\[
{\rm U}=\left(\begin{array}{ccc}
  \cos\psi & \sin\psi & 0 \\
  -\sin\psi & \cos\psi & 0 \\
  0 & 0 & 1 \\\end{array}\right)
\]
relates different orbits $\mathcal E_{a,b}$ and $\mathcal E_{\alpha^2a,\,\alpha b}$
of $e(3)$ and changes the form of original Hamiltonian by
\bq\label{Hrot}
H\to \widetilde{H}=J_1^2+J_2^2+ 2J_3^2+2\widetilde{c}_1x_1 +
2\widetilde{c}_2x_2\,,\qquad \widetilde{c}_1,\widetilde{c}_2\in \mathbb C.
\eq
Here $\widetilde{c}_1=\alpha\,c_1\,\cos(\psi)$ and
$\widetilde{c}_2=\alpha\,c_1\,\sin(\psi)$.
\par\noindent
\textbf{Example 2:} Transformation (\ref{shiftE3}) with
\[
{\rm S}=\left(\begin{array}{rrr}
  0 & 1 & 0 \\
  -1 & 0 & 0 \\
  0 & 0 & 0 \\\end{array}\right)
\]
changes the form of the  Hamiltonian (\ref{H-Kow}) as follows
$$
H\to{\widetilde H}=J_1^2+J_2^2+ 2J_3^2+2c_1
x_1+2 (x_2J_1-x_1J_2)+(x_1^2+x_2^2)\,.
$$
This form of the  Hamiltonian involves the third component of the vector
product $\mathbf J \times \mathbf x$ and at the first glance looks similar to
the deformed Hamilton function (\ref{sHam3}), which contains the first component.
However transformations (\ref{rotE3}) and (\ref{shiftE3}) are not
enough to relate the deformed Kowalevski top (\ref{sHam3})
and the original Kowalevski top (\ref{H-Kow}) on $e(3)$.

Let parameter $\alpha$ and matrices $U$ and $S$ in (\ref{rotE3}) and (\ref{shiftE3})
be functions of the Casimir elements $A,B$. In this
case the transformations  remain to be Poisson mappings. Such
Poisson maps change the form of the Hamiltonian as a function on the whole Poisson
manifold. For instance, the Hamilton function (\ref{Hrot}) becomes
\[
\widetilde{H}(A,B)=J_1^2+J_2^2+ 2J_3^2+2\widetilde{c}_1(A,B)\,x_1 +
2\widetilde{c}_2(A,B)\,x_2,
\]
where $\widetilde{c}_1(A,B)$ and $\widetilde{c}_2(A,B)$ are arbitrary functions
on the Casimir elements (\ref{caz0}). Of course, on each symplectic leaf the
function $\widetilde{H}(A,B)$ coincides with (\ref{Hrot}) and,
therefore, the above construction of nonlinear Poisson mappings is trivial.

\subsubsection{Generalized shifts of $\mathbf J$}
Consider the following generalizations of
transformations (\ref{shiftE3}):
\bq\label{autoE3}
\mathbf x\to \mathbf x \,,\qquad \mathbf J\to \mathbf J+\mathbf g(\mathbf x),
\eq
where components  $\textrm{g}_k(x_1,x_2,x_3)$ of the vector $\mathbf g$ are
nonlinear functions of the Poisson vector $\mathbf x$. Substituting
new variables into (\ref{e3}) we arrive at the following  conditions on the vector
$\mathbf g$:
\bq
{\rm div}\, \mathbf g=2 \beta'(A), \qquad (\mathbf x \cdot \mathbf
g)=\beta(A), \label{g-cond}
\eq
where $\beta$ is an arbitrary function of the Casimir element $A=\mathbf x^2$.

\begin{prop}
General solution of equations (\ref{g-cond}) is given by
$$
\mathbf g= \mathbf x \times ({\rm grad}\,W + F\,\mathbf n)+\beta\,\mathbf f,
$$
where potential $W({\mathbf x})$ is an arbitrary scalar function of $\,{\mathbf x}$,
$F$ is an arbitrary scalar function of two variables $~x_1+x_2+x_3,~$ and
$x_1^2+x_2^2+x_3^2$, and  vectors $\mathbf n$ and $\mathbf f$ are given by
\[\mathbf n=(1,1,1),\qquad \mathbf
f=\left(\displaystyle\frac{x_1}{x_1^2+x_2^2},\frac{x_2}{x_1^2+x_2^2},0\right).\]
Under the  transformation (\ref{autoE3}) the values of the Casimir
functions are changed as
$$
\widetilde{a}={a}\,,\qquad \widetilde{b}=b+\beta(a)\,.
$$
\end{prop}
Thus (\ref{autoE3}) is a nonlinear canonical transformation which relates
the symplectic manifolds $\mathcal E_{ab}$ and $\mathcal E_{\widetilde{ab}}$.
One can apply this transformation in order to get ``new'' integrable systems on
these manifolds.

\subsubsection{Generalized rotations}

Consider generalized rotations of the form
\[
\mathbf x\to \widetilde{\mathbf x}=\alpha(\mathbf x, \mathbf J)\,{\rm U}(\mathbf x, \mathbf
J)\, \mathbf x\,,\qquad \mathbf J\to\widetilde{\mathbf J}={\rm U}(\mathbf x, \mathbf
J)\,\mathbf J\,,
\]
where the scalar factor $\alpha(\mathbf x, \mathbf J)$ and the orthogonal matrix
${\rm U}(\mathbf x, \mathbf J)$ are some functions of variables $\mathbf x$ and $\mathbf J$.
Requiring this transformation to be a Poisson map,  we obtain a system of partial
differential equations for  $\alpha(\mathbf x,\mathbf J)$ and
${\rm U}_{ij}(\mathbf x, \mathbf J)$. It would be interesting to find a general solution
of this system.   Here we consider a particular case when $\alpha(\mathbf J)$
and  ${\rm U}_{ij}(\mathbf J)$ depend on one (say, third) component of angular momenta only.

\begin{prop}
Let  $f(J_3)$ and $g(J_3)$ be any functions such that
\[ f^2+g^2=c^2=const,\]
then the mapping
\bq \label{auto-e31}
\varphi:~\mathbf x\to \sqrt{f^2+g^2}\,{\rm U} \mathbf x,\qquad \mathbf J \to {\rm U} \mathbf
J\,,
\eq
where \[ {\rm U}=\dfrac{1}{\sqrt{f^2+g^2}}\left(\begin{array}{ccc}
  f & g & 0 \\
  -g & f & 0 \\
  0 & 0 & \sqrt{f^2+g^2}
\end{array}\right)\,
\]
is a canonical transformation of  $e(3)$, which changes the values of
Casimir functions (\ref{caz0}) by the rule
\[
{\widetilde a}=ac^2\,,\qquad  \widetilde{b}\, =bc\,.
\]
\end{prop}
The generalized rotation (\ref{auto-e31}) changes the form of the original
Hamiltonian for the Kowalev\-ski top (\ref{H-Kow}) as follows
$$
H\to\widetilde{H}= {J_1}^{2}+{J_2}^{2} + 2{J_3}^{2}
 +x_1\, f(J_3) + x_2\,g(J_3)\,.
$$
In particular, with the help of such transformation we can obtain
the following exotic Hamiltonian
\[
\widetilde{H}= {J_1}^{2}+{J_2}^{2} + 2{J_3}^{2}
 +x_1\, \sin(J_3) + x_1\,\cos(J_3).
\]

\subsection{Poisson map  between  $so(4)$ and $e(3)$ }

In this subsection we consider Poisson maps between Poisson manifolds of $so(4)$ with
generators ${\widetilde J_i}, y_j$ and $e(3)$  with
generators $J_i, x_j$.
We restrict ourselves to special maps of the form
$$
\widetilde {\mathbf J} = {\mathbf J},\qquad \mathbf y=\alpha(A,B)\,
\mathbf x+{\rm U}(x_1,x_2,x_3,A,B)\,\mathbf J,\,
$$
where $\alpha$ is a scalar function of Casimir elements (\ref{caz0}) and
$U$ is a matrix, which is not assumed to be orthogonal.
Such maps identify the rotation subalgebras of  $so(4)$ and $e(3)$.

The relations $\bigl\{J_i\,,y_j\,\bigr\}=\varepsilon_{ijk}y_k $ between components
of the vectors $\mathbf y$ and $\mathbf J$ bring to an overdetermined
system of partial differential equations for the matrix ${\rm U}$.
This system has the following general solution
\[{\rm U}=\beta(A,B)\, \mbox{I}_{3}+ \gamma(A,B)
\left(
\begin{array}{ccc}
  0 & x_3 & -x_2 \\
  -x_3 & 0 & x_1 \\
  x_2 & -x_1 & 0 \\
\end{array}
\right)+\delta(A,B)\left(\begin{array}{ccc}
  x_1^2 & x_1x_2 & x_1x_3 \\
  x_1x_2 & x_2^2 & x_2x_3 \\
  x_1x_3 & x_2x_3 & x_3^2 \\
\end{array}\right)
\]
depending on arbitrary functions $\beta,\gamma$ and $\delta$ of the Casimir elements
(\ref{caz0}), where $\mbox{I}_{3}$ is a unit $3\times 3 $ matrix.
Notice that if  $\beta+A\delta =1,\,$ $\gamma^2=(\beta+1)\delta,$
the above formula for $U$ coincides with the well-known
Gibbs representation  \cite{gibbs} of an arbitrary orthogonal matrix.

The relations $ \bigl\{ y_i\,,y_j\,\bigr\}= \varepsilon_{ijk}\varkappa^2\,J_k $ give
rise to a system of algebraic equations for $\alpha, \beta,\gamma$
and $\vk,$ which has only
two different solutions. In the first case  $\gamma=0,\,
\beta^2=\varkappa^2,\,\alpha=-A\,\delta$ and the solution describes a
reduction $e(3)$ to $so(3)$ by the trivial scaling  $\mathbf y=\vk\mathbf J$.

The second solution is:
\[\beta=0,\qquad\gamma^2=-\dfrac{\varkappa^2}{A}
\]
 and $\delta$ is
arbitrary function which can be removed by shift $\alpha \rightarrow \alpha+A\,\delta$.
Thus this solution corresponds to the transformation
\begin{equation} \label{rTrans}
\zeta:\qquad\mathbf J\to\mathbf J,\qquad \mathbf y= \alpha~\mathbf x+
\gamma~\mathbf x \times \mathbf J,
\end{equation}
which maps the manifold $e(3)$ to the manifold $so(4)$. Below we consider a special case
$\alpha=const$ in more detail.
\begin{prop}
Suppose  $\alpha\neq 0$ is a constant and $\gamma$ is a solution of equation
\bq \label{Ceq1}
 A\gamma^2+\vk^2=0\,.
\eq
Then transformation (\ref{rTrans}) is a Poisson map of $e(3)$ to $so(4)$.

The inverse Poisson map $so(4)\rightarrow e(3)$ is given by
\begin{equation} \label{irTrans}
{\mathbf x}= \frac{\alpha^2\,\mathbf y+\gamma_{\vk}^2\,(\mathbf y
\cdot \mathbf J)\,\mathbf J+\alpha\,\gamma_{\vk}\,(\mathbf y\times
\mathbf J)}
  {\alpha\,(\,\alpha^2 +\gamma_{\vk}^2\mathbf J^2\,)},
\end{equation}
where the algebraic function $\gamma_{\vk}(A_{\vk}, B_{\vk})$ depending on
$so(4)$-Casimir elements (\ref{center}) is defined  by
\bq \label{cond}
 B_{\vk}^2\,\gamma_{\vk}^4+A_\vk\alpha^2\gamma_{\vk}^2+\alpha^4\vk^2=0\,.
\eq
Notice that the branches of square roots in
(\ref{Ceq1}) and (\ref{cond}) have to be consistent.

The Poisson maps (\ref{rTrans}) and (\ref{irTrans})
give rise to the symplectic correspondence between the symplectic submanifolds ${\cal
E}_{ab}$ in $e(3)$ and symplectic submanifolds ${\cal O}_{a'b{\,}' }$ in $so(4)$, where
\begin{equation} \label{rCaz1}
a'=\alpha^2\,a+\frac{\vk^2\,b^2}{a},\qquad b{\,}'=\alpha\,b.
\end{equation}
\end{prop}

Obviously, compositions of the Poisson maps (\ref{rTrans}) and (\ref{irTrans}) with
canonical transformations of $e(3)$ or $so(4)$ give rise to different Poisson maps
relating $e(3)$ and $so(4)$.

The singular points of the transformation can be easily seen from the formulas
(\ref{rTrans})-(\ref{irTrans}).

It turns out that the Poisson maps (\ref{rTrans}) and (\ref{irTrans}) establish a
correspondence between the reduced four-dimensional phase spaces of the Kowalevski
top on $so(4)$ and the deformed Kowalevski top on $e(3):$
\begin{theor}
Transformation (\ref{rTrans}) sends
the Hamilton function
$$
H_\vk=J_1^2+J_2^2+ 2J_3^2+2\widetilde{c}_1 y_1\,
$$
on $\mathcal O_{a'b'}$ to the Hamilton function
\bq \label{sHam33}
\widehat{H}=J_1^2+J_2^2+2J_3^2 +2{c}_1x_1+2{c}_2(x_2J_3-x_3J_2)
\eq
on $\mathcal E_{ab}$, where $c_1=\alpha\,\widetilde{c}_1,$
$c_2=\gamma\,\widetilde{c}_1$ and the constants $a,b,a',b'$ are related by
(\ref{rCaz1}).
\end{theor}

Notice that $c_{1}$ in formula (\ref{sHam33}) is a constant whereas $c_{2}$
is a function of the Casimir element $A$. However, on each symplectic leaf
$c_{1}$ and $c_{2}$ are constants and $\widehat{H}$ from (\ref{sHam33})
coincides with (\ref{sHam3}).

{\bf Remark.} In \cite{bormam1} a different Poisson map
$$
\mathbf J\to \mathbf J, \qquad \mathbf y\to \mathbf x=
\frac{\mathbf J\times (\mathbf y \times \mathbf J)}
{\vert \mathbf J\times (\mathbf y \times \mathbf J)\vert}
$$
from $so(4)$ to $e(3)$ was considered. This mapping takes any
symplectic leaf $\mathcal O_{a'b'}$ of $so(4)$
to the same symplectic leaf $(\mathbf x, \mathbf J)=0,\, \mathbf x^{2}=1$
of $e(3)$ and therefore it is not invertible. This mapping allows to lift
integrable Hamiltonians from $e(3)$ to $so(4)$ but it involves
radicals and don't preserve the property of the Hamiltonians to be rational.

\section{Lax representation for the so(4) Kowalevski top}
\setcounter{equation}{0}

A Lax representation
\bq\label{Lax-Eq}
\frac{d}{dt}{ L}=[M, L]
\eq
for the Kowalevski top (\ref{H-Kow}) was found by Reyman and
Semenov-Tian-Shansky \cite{rs87}. The corresponding Lax matrices are
\ben \label{Lax-Kow}
L(\lambda)
&=& \left(\begin{array}{ccccc} 0& J_3& -J_2& \lambda & 0\\-J_3& 0& J_1& 0&\lambda\\
J_2& -J_1& 0&0&0\\ \lambda&0& 0& 0& -J_3\\0&\lambda&0& J_3&
0\end{array}\right)- \dfrac{c_1}{\lambda}
\left(\begin{array}{ccccc}0&0&0&x_1&0
                        \\0&0&0&x_2&0
                        \\0&0&0&x_3&0
                        \\x_1&x_2&x_3&0&0
                        \\0&0&0&0&0\end{array}\right)
\\
&\defeq&\lambda\, {\mathcal A}+\sum_{i=1}^3 J_i\cdot{\mathcal
J}_i-\dfrac{c_1}{\lambda}\,\sum_{i=1}^3x_i\cdot{\mathcal X}_i
\nn
\en
and
\ben \label{M-Kow}
M(\lambda)=2
\left(\begin{array}{ccccc} 0& -2J_3& J_2& -\lambda& 0\\
2 J_3& 0& -J_1& 0&-\lambda\\
-J_2&J_1& 0&0&0\\-\lambda&0& 0& 0&0\\0&-\lambda&0& 0&
0\end{array}\right).
\en
The characteristic curve ${\rm Det}(L(\lambda)-\mu\cdot \mbox{\rm I})=0$,
where $\mbox{\rm I}=diag(1,1,1,1,1)$ is the unit matrix, provides a
complete set of first integrals for the Kowalevski top \cite{brs89}.

It is essential for general group-theoretical approach to integrable
systems \cite{rs87} that the matrices $\mathcal A,~ \mathcal
J_i,~\mathcal X_i$  belong to the matrix realization of the Lie
algebra $so(3,2)$ by $5\times5$ matrices $\mathcal Z$ satisfying the
identity
\bq \label{so32}
\mathcal Z^T=-\mbox{\rm I}_{3,2}\mathcal Z \mbox{\rm I}_{3,2},
\eq
where $\mbox{\rm I}_{3,2}=diag(1,1,1,-1,-1)$. The Lax matrices
(\ref{Lax-Kow}) are invariant with respect to the following
involution
$$
\tau:\qquad Z(\lambda)\to -Z^T(-\lambda).
$$
Using the well-known isomorphism $so(3,2)\simeq sp(4,\mathbb R)$ one
can obtain also a $4\times 4$ Lax pair for the Kowalevski top
\cite{brs89}.

A Lax representation for the deformed Kowalevski top on $e(3)$ with
the Hamilton function (\ref{sHam3}) was found in \cite{sokts1}. This
representation involves an additional matrix
$$
Y=\left(\begin{array}{ccccc}
0&0&0&x_1&0\\
0&0&0&x_2&0\\
0&0&0&x_3&0\\
-x_1&-x_2&-x_3& 0& 0\\
0&0& 0& 0&0
\end{array}\right)
 \defeq
 \sum_{i=1}^3 x_i\mathcal Y_i.
$$
The constant matrices $\mathcal Y_i$ are
symmetrized anticommutators of matrix coefficients of the initial
Lax matrix $L$ (\ref{Lax-Kow})
\[\mathcal Y_i=\varepsilon_{ijk}\left(\mathcal X_j
\mathcal J_k+ \mathcal J_k\mathcal X_j\right).
\]
They do not respect  involution (\ref{so32}) and hence do not
belong to the algebra $so(3,2)$.
\begin{prop}{(\rm Sokolov, Tsiganov \cite{sokts1})}
The flow with the Hamiltonian $\widehat{H}$ (\ref{sHam3}) is
equivalent to the matrix differential equations
\bq \label{laxtr}
\dfrac{d}{dt}\widehat{L}_i(\lambda)=\widehat{L}_i(\lambda)\,\widehat{M}(\lambda)+
\widehat{M}\,^T(-\lambda)\,\widehat{L}_i(\lambda), \qquad i=1,~ 2,
\eq
where
\bq \label{kL12}
\widehat{L}_1(\lambda)=L(\lambda)+\frac{c_2}2\sum_{i=1}^3x_i\cdot\Bigl((\mathcal
X_i-\mathcal Y_i)\mathcal A-\mathcal A(\mathcal X_i+\mathcal
Y_i)\Bigr)\,,\qquad \widehat{L}_2(\lambda)=-\mbox{\rm
I}+\frac{c_2}{\lambda}Y,
\eq
$$
\widehat{M}=M+2c_2\left( \begin{array}{ccccc}
 x_1&0& 0& 0& 0\\  x_2&0& 0& 0& 0\\
 x_3&0& 0& 0& 0\\
  0& 0& 0& -x_1&0\\
  0& 0& 0& -x_2&0
\end{array}
\right) \,,
$$
and the superscript $T$  stands for matrix transposition.
\end{prop}

It is easy to verify that the matrices $\widehat L_{1,2}$ (\ref{kL12})
can be rewritten as follows
\bq\label{tr-lax}
\widehat{L}_1=(\mbox{\rm I}-g^\tau)^{-1}\,L\,(\mbox{\rm I}-g)+V,\qquad
\widehat{L}_2=-(\mbox{\rm I}-g^\tau\Bigr)^{-1}\,(\mbox{\rm
I}+g^\tau\,g)\,(\mbox{\rm I}-g),
\eq
where
\[g=\dfrac{c_2}{\lambda}\left(\begin{array}{ccccc}
0& 0& 0& x_1&0\\
0&0& 0&x_2&0\\0&0&0& x_3&0\\0&0& 0& 0&0\\0&0&0& 0&
0\end{array}\right),
\]
and
\[
V=-\dfrac{c_2}{\lambda}\left(\begin{array}{ccccc}
0& 0& 0& (\mathbf x\times\mathbf J)_1&0\\
0&0& 0& (\mathbf x\times\mathbf J)_2&0\\0&0&0& (\mathbf x\times\mathbf J)_3&0\\
(\mathbf x\times\mathbf J)_1&(\mathbf x\times\mathbf J)_2& (\mathbf
x\times\mathbf J)_3& 0&0\\0&0&0& 0& 0\end{array}\right)\,.
\]
Notice that the matrix $V$ depends on the components $({\mathbf
x}\times \mathbf J)_{i}$ of the cross product ${\mathbf x}\times
\mathbf J$ only.

Relations (\ref{laxtr}) imply that matrices
\begin{equation} \label{newlax2}
\widehat{L}_+=\widehat{L}_1(\lambda)\,\widehat{L}_2^{-1}(\lambda), \qquad
\widehat{L}_-=\widehat{L}_2^{-1}(\lambda)\,\widehat{L}_1(\lambda)
\end{equation}
satisfy  the usual Lax equations (\ref{Lax-Eq})
$$
\frac{d}{dt} \widehat{L}_+=\left[ \widehat{L}_+,\,
-\widehat{M}\,^T(-\lambda)\right], \qquad \frac{d}{dt} \widehat{L}_-=\left[
\widehat{L}_-,\,\widehat{M}(\lambda) \right].
$$

The explicit form of the Lax matrices  (\ref{newlax2}) is rather
complicated. However matrices $\widehat L_\pm$ can be simplified with the
help of a gauge transformation. Let us define a new matrix
$\widehat{L}$ by the formula
$$
 \widehat{L}(\lambda)=-(\mbox{\rm I}-g)\,\widehat{L}_-\,(\mbox{\rm I}-g)^{-1}.
$$
Using (\ref{tr-lax}) and the following property of $V$:
\[
(\mbox{\rm I}-g^{\tau})\,V\,(\mbox{\rm I}-g)^{-1}=V,
\]
this matrix can be rewritten in the form
\[\widehat{L}(\lambda)=(\mbox{\rm
I}+g^\tau\,g)^{-1}\,\Bigl(\,L(\lambda)+V\,\Bigr).
\]
It can be verified that
$$
 \widehat{L}^\tau(\lambda)=-(\mbox{\rm
I}-g^\tau)\,\widehat{L}_+\,(\mbox{\rm I}-g^\tau)^{-1}\,.
$$

The next statement describes a Lax pair for the deformed Kowalevski
top on $e(3)$ with Hamiltonian (\ref{sHam3}) related to the matrix
$\widehat{L}$.
\begin{prop}
The flow with the Hamiltonian $\widehat{H}$ (\ref{sHam3}) is
equivalent to the Lax equation (\ref{Lax-Eq}), where
\bq\label{Lax-def1}
\widehat{L}(\lambda)= (\mbox{\rm
I}+g^\tau\,g)^{-1}\,\Bigl(\,L(\lambda)+V\,\Bigr)\qquad\mbox{\rm
and}\qquad \widehat{M}(\lambda)=M(\lambda)\,(\mbox{\rm I}+g^\tau\,g).
\eq
\end{prop}
It is important that the product $ g^\tau\,g$ depends on the Casimir
function only:
\bq
\label{defG}
 g^\tau\,g=\dfrac{c_2^2\mathbf x^2}{\lambda^2}\,\mathcal G,\qquad
\mathcal G=\left(\begin{array}{ccccc}
0& 0& 0& 0&0\\
0&0& 0& 0&0\\0&0&0& 0&0\\0&0& 0& 1&0\\0&0&0& 0&0\end{array}\right).
\eq
Taking into account this formula, we get
$$
\widehat{L}(\lambda)= \left(\mbox{\rm
I}-\dfrac{c_2^2\mathbf x^2}{\lambda^2+c_2^2\mathbf x^2}\,\mathcal
G\right) \left(\lambda\, {\mathcal A}+\sum_{i=1}^3 J_i\cdot{\mathcal
J}_i-\dfrac{1}{\lambda}\,\sum_{i=1}^3y_i\cdot{\mathcal X}_i\right),
$$
where $y_i=c_1x_i+c_2(\mathbf x\times \mathbf J)_i $.
We see that in the case $c_{2}=0$ the matrix $\widehat{L}$ coincides
with (\ref{Lax-Kow}).

Thus in order to construct the Lax matrices (\ref{Lax-def1}) for the
deformed Kowalevski top on $e(3)$ we have to substitute
$y_i$ instead of $x_i$ into
the Lax matrices found by Reyman and Semenov-Tian-Shansky \cite{rs87} and
multiply the result by matrices depending on the Casimir element
only.

The fact that $\widehat{L}$ depends only on variables $J_{i}$ and
$y_i=c_1x_i+c_2(\mathbf x\times \mathbf J)_i$, which define
transformation (\ref{rTrans}), allows us to construct a Lax
representation for the Kowalevski top on $so(4).$ Namely, an obvious
combination of Proposition 3 and Proposition 5 leads to
\begin{theor}
The matrices
\ben
\label{L-o4}
L_\vk(\lambda)&=&\left(\mbox{\rm
I}+\dfrac{\widetilde{c}_1^2\vk^2}{\lambda^2-\widetilde{c}_1^2\vk^2}\,\mathcal
G\right)
\times\nn\\
\\
&\times& \left[
\left(\begin{array}{ccccc} 0& J_3& -J_2& \lambda & 0\\-J_3& 0& J_1& 0&\lambda\\
J_2& -J_1& 0&0&0\\ \lambda&0& 0& 0& -J_3\\0&\lambda&0& J_3&
0\end{array}\right)- \dfrac{\widetilde{c}_1}{\lambda}
\left(\begin{array}{ccccc}0&0&0&y_1&0\\0&0&0&y_2&0\\0&0&0&y_3&0
\\y_1&y_2&y_3&0&0\\0&0&0&0&0\end{array}\right)\right] \nn
\en
and
\ben
\label{M-o4}
M_\vk=2\left(\begin{array}{ccccc} 0& -2J_3&J_2& -\lambda& 0\\
2 J_3& 0& -J_1& 0&-\lambda\\
-J_2& J_1& 0&0&0\\-\lambda&0& 0& 0&0\\0&-\lambda&0& 0&
0\end{array}\right)\left(\mbox{\rm
I}-\dfrac{\widetilde{c}_1^2\vk^2}{\lambda^2}\mathcal G\right),
\en
where $\mathcal G$ is given by (\ref{defG}),
define a Lax
representation for the $so(4)$-Kowalevski top with the Hamiltonian
\[H_\vk=J_1^2+J_2^2+ 2J_3^2+2\widetilde{c}_1 y_1,
\qquad \widetilde{c}_1\in \mathbb C.\]

The  characteristic curve ${\rm Det}(L_\vk(\lambda)-\mu \mbox{\rm
I})=0$ provides a complete set of first integrals of motion
$$
(\widetilde{c}_1^2\vk^2-\lambda^2)\mu^4
+\mu^2(2\lambda^4-(H_\vk+\widetilde{c}_1^2\vk^2)\lambda^2+\widetilde{c}_1^2A_\vk)=
\lambda^6 -H_\vk\lambda^4-K_\vk\lambda^2-\widetilde{c}_1^2B_\vk^2\,.
$$
\end{theor}
It is seen that the Lax pair (\ref{L-o4}), (\ref{M-o4}) on $so(4)$ is
formed from the  Lax pair (\ref{Lax-Kow}), (\ref{M-Kow}) on $e(3)$
by substitution $x_{i} \to y_{i}, \quad c \to \tilde c$
and multiplication by the $\lambda$-meromorphic diagonal constant matrix
factors from the left and right correspondingly
\ben
L_\vk(\lambda)&=& \left(\mbox{\rm
I}+\dfrac{\widetilde{c}_1^2\vk^2}{\lambda^2-\widetilde{c}_1^2\vk^2}\,\mathcal
G\right) \left.L(\lambda)\right |_{x_{i} \to y_{i},~  c_1 \to \tilde c_1},
\nn\\
\label{Lax-o44}\\
M_\vk(\lambda)&=&M(\lambda)\vert_{c_1 \to \tilde
c_1}\left(\mbox{\rm
I}-\dfrac{\widetilde{c}_1^2\vk^2}{\lambda^2}\mathcal G\right).\nn
\en This Lax pair for the Kowalevski top on $so(4)$ allows us to
apply the standard finite-band integration technique to this
system.

The following comments are in order:
\begin{itemize}
\item
 Multiplying $L_{\vk}$  by the factor
$\lambda^{2}-\widetilde{c}_1^{2} \vk^{2},$ one can remove the
poles at $\lambda=\pm \widetilde{c}_1 \vk$. Nevertheless just
operator $L_{\vk}$ tends to the original Lax matrix from
\cite{rs87} as $\vk \rightarrow 0$. Probably this means that the
poles in the Lax matrix for the Kowalevski top on $so(4)$ are
essential.

\item
 Substituting the Lax matrices for the Kowalevski gyrostat on
$e(3)$ (see \cite{rs87}) for $L$ and $M$ in (\ref{Lax-o44}), one
gets a Lax pair for the Kowalevski gyrostat on $so(4)$.

\item The matrices $\widehat{L}_{1,2}$  and $\widehat{L}$  do not
respect the involution (\ref{so32}) and, therefore,  are out
of the matrix realization of the Lie algebra $so(3,2)$.
 They can not be rewritten as $4\times 4$ matrices
via the isomorphism $so(3,2)\simeq sp(4)$. Nevertheless precisely
the matrices $\widehat{L}_{1,2}$  and $\widehat{L}$ provide a
multi-dimensional generalization of the Kowalevski gyrostat
\cite{sokts1}.
\end{itemize}

Applying the  Poisson map  from Proposition 3 to a similar Lax
matrix for the Lagrange top on $e(3)$  \cite{rs87}, we
get the following
\begin{prop}
For the Lagrange top on $so(4)$ defined by the Hamiltonian
\[H^{Lag}_\varkappa=J_1^2+J_2^2+ J_3^2+2 c\, y_1,\qquad
c \in \mathbb C
\]
a Lax matrix is given by
\[
L_\varkappa^{Lag}(\lambda)=\left(\begin{array}{cccc}
1& 0& 0& 0\\[2mm]0& 1& 0& 0\\[2mm]0& 0& 1& 0\\[2mm]
0& 0& 0&
1+\frac{\dsize c^2\varkappa^2}{\dsize \lambda^2-c^2\varkappa^2}\end{array}\right)
\left(\begin{array}{cccr}
0& J_3& -J_2& \lambda-\frac{\dsize c y_1}{\dsize \lambda}\\[2mm]
-J_3& 0& J_1& -\frac{\dsize c y_2}{\dsize \lambda}\\[2mm]
J_2& -J_1& 0& -\frac{\dsize c y_3}{\dsize \lambda}\\[2mm]
\lambda-\frac{\dsize c y_1}{\dsize\lambda}&-\frac{\dsize c y_2}{\dsize \lambda}&
-\frac{\dsize c y_3}{\dsize \lambda}& 0
\end{array}\right).
\]
The corresponding characteristic curve
\[
(c ^2\varkappa^2-\lambda^2)\mu^4
-(\lambda^4-H_\varkappa^{Lag}\lambda^2+c^2 A_\varkappa) \mu^2=
\left(K_\varkappa^{Lag}\lambda^2-c B_\varkappa\right)^2,
\]
where $K_\varkappa^{Lag}=J_1$, provides a complete set of
integrals of motion.
\end{prop}

\section{Separation of variables}
\setcounter{equation}{0}
The separation of variables for the Hamilton function $H_\vk$ (\ref{Hk}) on
$so(4)$ was obtained in \cite{komkuz} by a non-canonical reduction to the
Neumann system. (Beforehand in the unpublished calculations the first author
obtained the result by a variant of the original  Kowalevski approach.)
The results of \cite{komkuz} were based on the fact that the evolutionary
equations for the Kowalevski top, written in special variables of Haine
and Horozov \cite{HH}, coincides with the Neumann system.

Applying Propositions 1 and 2 one can derive explicit formulas for
separation of variables for the model (\ref{sHam3}) on $e(3)$
from the paper \cite{komkuz}. But historically we have obtained these
formulas following the original Kowalevski work \cite{kow89}.
Below we follows this line.

Consider the Hamiltonian on $e(3)$
$$
\widehat{H}=J_1^2+J_2^2+2J_3^2 +2{c}_1x_1+2{c}_2(x_2J_3-x_3J_2).
\eqno{(2.13)}
$$
It is easy to prove that  variables
$$
 z_1=J_1+iJ_2, \qquad z_2=J_1-iJ_2,
$$
satisfy the following system of equations
$$
{\dot{z}_1}^2-F(z_1)+\widehat{\xi}\,(z_1-z_2)^2=0\,,\qquad
{\dot{z}_2}^2-F(z_2)+\widehat{\xi}^*\,(z_1-z_2)^2=0\,
$$
where $\widehat{\xi},~ \widehat{\xi}^*$ are given by (\ref{xi})
with  $\vk=0$:
$$
\widehat{\xi}=\xi-c_2\{\mathbf J^2,x_1+ix_2\}-c_2^2A, \qquad
 \widehat{\xi}^*=\xi^*-c_2\{\mathbf J^2,x_1-ix_2\}
-c_2^2 A.
$$
Here $F(z)$ is a polynomial of four degree with coefficients being integrals of
motion
$$
F(z)=z^4-2\widehat{H}\,z^2+8c_1B\,z +\widehat{K}
-4A\,c_1^2+2c_2^2\,(2B^2-\widehat{H}\,A)-c_2^4\,A^2.
$$
According to \cite{kow89}, we define the biquadratic form
\[
F(z_1,z_2)=\dfrac12\Bigl(F(z_1)+F(z_2)-(z_1^2-z_2^2)^2\Bigr)
\]
and the separated variables
\bq
s_{1,2}=\dfrac{F(z_1,z_2)\pm\sqrt{F(z_1)\,F(z_2)\,}}{2(z_1-z_2)^2} \label{SepVar}
\eq
such that
\bq
\label{s-eqs} \dot{s}_1=\dfrac{\sqrt{P_5(s_1)\,}}{s_1-s_2}\,,\qquad
\dot{s}_2=\dfrac{\sqrt{P_5(s_2)\,}}{s_2-s_1}\,,\qquad P_5(s)=P_3(s)P_2(s)\,.
\eq
Here $P_3(s)$ and $P_2(s)$ are polynomials of third and second degree:
\ben
P_3(s)&=&s\Bigl(4s^2+4s\,\widehat{H}+\widehat{H}^2-\widehat{K}+4c_1^2A
+2c_2^2\,(\widehat{H}\,A-2\,B^2)+c_2^4A^2\Bigr)+4c_1^2\,B^2,  \nn\\
\nn\\
P_2(s)&=&4s^2+4(\widehat{H}+c_2^2\,A)\,s+\widehat{H}^2-\widehat{K}+2c_2^2\,\widehat{H}\,A
+c_2^4\,A^2.\nn
\en
To integrate  equations  (\ref{s-eqs}) one should substitute the values of
integrals of motion and Casimir elements found from initial data. Equations
(\ref{s-eqs}) are integrated in terms of genus two hyperelliptic functions of
time.

As well as in the case of initial Kowalevski top \cite{nv85} one can check by
direct computations that functions  $s_{1,2}$ (\ref{SepVar}) defined on the
whole phase space commute with respect to initial Poisson brackets (\ref{e3})
$$
\{s_1,s_2\}=0 .
$$
The reasons why the functions $s_1, s_{2}$ give rise to canonical
variables on $e(3)$ seem to be unclear (see comments in \cite{nv85}, where
the Poisson commutativity of $s_1$ and $s_2$ was originally pointed out).

The  momenta $p_{1,2}$ conjugated to coordinates $s_{1,2}$ can be introduced
according to \cite{nv85} (see  \cite{komkuz} for another approach).
The result is
\bq
  p_{i}=\dfrac{1}{4\sqrt{s_{i}}}\,\ln\left(
\dfrac{2\sqrt{s_{i}P_5(s_{i})\,}-P_3(s_{i})-s_{i}P_2(s_{i})}{4(as_{i}+b^2)
(c_1^2-c_2^2 s_{i})}
  \right)\,.
  \label{pS}
\eq
In variables (\ref{SepVar}), (\ref{pS}) the Hamilton function (\ref{sHam3}) is
given by
\[
\widehat{H}=-s_1-s_2+\dfrac{c_1^2b^2}{2s_1s_2}-\dfrac{a\,c_2^2}2
+\dfrac{d_1\cosh(4p_1\sqrt{s_1})-d_2\cosh(4p_2\sqrt{s_2})}{2(s_1-s_2)},
\]
where
\[d_i=\dfrac{(c_2^2\,s_i-c_1^2)(a\,s_i+b^2)}{s_i}.\]
Using this relation, we obtain two separated equations
\bq \label{SepEqHd}
  2s_i^3+\left(2\widehat{H}+c_2^2\,a\right)s_i^2-\kappa\,s_i+c_1^2b^2=
(c_2^2\,s_i-c_1^2)(a\,s_i+b^2)\cosh(4p_i\sqrt{s_i}),
\eq
where
$$
4\kappa=(\widehat{H}+c_2^2\,a)^2-\widehat{K}+2c_1^2\,a\,.
$$
As usual, canonical variables $s_i$ (\ref{SepVar}) and $p_i$ (\ref{pS}) are
defined up to arbitrary canonical transformations that mix together $s_i$
and $p_i$ with the same $i$.  Evidently, such transformations
change the form of separated equations.
Equations (\ref{SepEqHd}) coincide with the separated equations from
\cite{komkuz} up to a canonical scaling of $p_i$ and $s_i$.
The mapping from \cite{komkuz} between the $so(4,\C)$ Kowalevski top and
the Neumann system relates the
separated  variables $s_1,s_2$ of Kowalevski top and separated variables
$\lambda_{1,2}$ for the Neumann top by $s_{1,2}=2\lambda_{1,2}+H$.

\section{Summary}
We present a Poisson map which relates rank-two Poisson manifolds $e(3)$
and $so(4)$. Using such transformations in rigid body dynamics we get new
results on deformations of the $e(3)$ and $so(4)$ Kowalevski tops.

The same Poisson map can be applied to another integrable systems,
for instance to the deformation of the Goryachev-Chaplygin gyrostat
proposed in \cite{sokts1}. In this case mapping (\ref{rTrans})-(\ref{irTrans})
sends the integrable Hamilton function on $e(3)$
\cite{sokts1}
\[ H^{g}=J_1^2+J_2^2+4J_3^2
+2\alpha{c}_1x_1+2\gamma{c}_1(x_2J_3-2x_3J_2),\]
to the following function on $so(4)$ manifold
\[ H_{\varkappa}^g=J_1^2+J_2^2+
4J_3^2-2c_1 y_1+\dfrac{2\vk^2 c_1J_3}{\gamma\mathbf y^2}\Bigl(
{\alpha}\,{y_2}+\gamma(y_1J_3-y_3J_1)\Bigr),
\]
where $A_\vk\gamma^2+\alpha^2\vk^2=0$.
This Hamiltonian commutes with
\[
K_\vk^g=c_1y_3J_1+\left(J_1^2+J_2^2-\dfrac{\vk^2c_1J_2}
{\gamma\mathbf y^2}\Bigl(\alpha
y_3-\gamma(y_1J_2-y_2J_1)\Bigr)\right)J_3
\]
on a special level of the Casimir
function $B_\vk=0$. Another version of the  Goryachev-Chaplygin top on
$so(4)$ was proposed in \cite{bormam1}.

One of our main  results is a Lax representation for the  Kowalevski  top on $so(4)$ provided by
Theorem 2. The Lax matrix $L_\vk(\lambda)$ (\ref{Lax-o44})  generate algebraic curve
different from the original Kowalevski one. Matrix $L_\vk(\lambda)$ should  originate
separated variables which in turn differ from that considered in section 5.
Such separation of variables remains an open question as well
as for the original $e(3)$  Kowalevski case.

\vskip.3cm
\noindent
{\bf Acknowledgments.}  The authors are grateful to M.A. Semenov-{T}ian-{S}hansky
for useful discussions. The research was partially supported by RFBR grants
02-01-00888 and 02-01-00431.

\end{document}